\documentclass[12pt,twoside]{article}
\usepackage{correl12}

\def\TheTitle{Strongly correlated electrons:   \newline Estimates of model parameters}
\def\TheHeading{Model parameters}                        
\def\TheAuthor{Olle Gunnarsson}                          
\def\TheInstitute{Max-Planck-Institut f\"ur Festk\"orperforschung}  
\def\TheAddress{Heisenbergstrasse 1, D-70569 Stuttgart, Germany}         
\def\TheChapter{5}                                       

\begin{document}
\MakeTitle           
\tableofcontents     
\newpage
\section{Introduction}

For weakly or moderately strongly correlated systems {\it ab initio} methods, such as
the density functional formalism\cite{Hohenberg,Kohn} or the GW method\cite{Hedin,FerdiGW}, 
are often quite successful. For strongly correlated systems, however, these methods
are often not sufficient. It is then necessary to treat correlation effects in a 
more accurate way. Such systems are often quite complicated with large unit cells.
It is then very hard to treat correlation effects within an {\it ab initio} approach, 
and one often turns to model Hamiltonians. The idea is then to focus on states and
interactions believed to be particularly important for the physics of interest. 
This has the additional advantage that it may then be easier to understand the       
physics, since less important effects do not confuse the interpretation. On the 
other hand, there is a risk of oversimplifying the model and thereby missing the 
correct physics. The purpose of this paper is to discuss this approach.

In principle it is straightforward to construct a model. We can produce a complete basis
set and then calculate matrix elements of the real space Hamiltonian
\begin{equation}
H=\sum_i\lbrack -{\hbar^2\over 2m}\bigtriangledown_i^2+
V_{ext}({\bf r}_i) \rbrack + \sum_{i<j}{e^2\over |{\bf r}_i-{\bf r}_j|}.
\end{equation}
For atoms or small molecules, this Hamiltonian may then be solved using
various many-body methods, e.g., configuration interaction (CI),
where the many-body wave function is written as a linear combination of 
determinants. For strongly correlated solids, however, a Hamiltonian
obtained in this way is often too complicated to allow reasonably accurate
calculations. We are then forced to use substantially simpler models.
This usually involves a drastic reduction of the basis set and the 
neglect of many interactions. Typical examples are the Anderson\cite{Anderson}, 
the Hubbard\cite{Hubbard} and the $t-J$\cite{tJ} models.

This approach involves the neglect of interactions which are large.
For instance, the Anderson impurity model is often used for a $3d$ 
impurity in a weakly correlated host.
We define a direct Coulomb integral
\begin{equation}\label{eq:1.2}
F_{ij}=e^2\int d^3r\int d^3r^{'} {\Phi_i^2({\bf r})
\Phi_j^2({\bf r}^{'})\over |{\bf r}-{\bf r}^{'}|},
\end{equation}
where $\Phi_i({\bf r})$ is the wave function of a state $i$.
Then the Coulomb integral $F_{3d,3d}$ between $3d$ electrons 
is kept, while, for instance, the integral $F_{3d,4s}$ between 
a $3d$ and a $4s$ electron is neglected. For a free Mn atom
$F_{3d,3d}=21$ eV and $F_{3d,4s}=10$ eV. Such an approximation is 
clearly highly questionable. An essential task is then to try to 
include explicitly neglected interactions or states implicitly as a 
renormalization of parameters\index{renormalization of parameters} in the model. As we show later, this leads 
to an effective Coulomb interaction between the $3d$ electrons which is
much smaller than the calculated value for a free atom. A basic assumption 
of such simple models is then that all the neglected interactions can 
with a reasonable accuracy be included implicitly as a renormalization 
of various model parameters\index{model parameters}. In this approach it is important to 
keep track of what effects are explicitly included in the model. These
should not be included in the calculation of parameters, since this
would involve double-counting. 

There are various ways of obtaining parameters. One approach has been indicated 
above. We use {\it ab initio} calculations to calculate parameters and then
we try to estimate how these are renormalized by neglected interactions.
Another is to calculate certain properties of the model, compare with experiment
and then adjust parameters until the experimental value is obtained. This 
approach then automatically gives renormalized parameters. It is important to try 
to obtain as much independent information as possible about the parameters, 
both from calculations and from different experiments, and to check if various 
pieces of information are consistent.

\begin{figure}[h]
 \centering
 \includegraphics[width=0.4\textwidth]{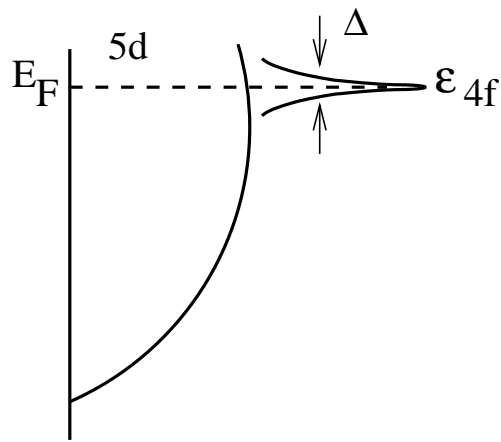}
 \caption{\label{fig:ce}Schematic density of state for a Ce compound according to the promotional model.}
\end{figure}

The importance of obtaining theoretical information about parameters can be 
illustrated by the historical development of the theory of Ce compounds.
Traditionally, Ce compounds were described in the so called promotional
model\cite{promotional}. It was assumed that the Ce $4f$ level was located very close to
the Fermi level, $E_F$, and that it had a very weak interaction with
other states. A mean-field theory was then used to show that this leads to 
a very narrow resonance, as indicated in Fig.~\ref{fig:ce}. The narrowness 
of the resonance could explain the large susceptibility and specific heat 
of Ce compounds, and the closeness of the $4f$ level to $E_F$ the change 
of apparent valence when the pressure or temperature were changed. Thermodynamic 
considerations, however, showed that the $4f$ level ought to be about 2 eV 
below $E_F$ \cite{Johansson}, in strong disagreement with the model. This 
result was later reconciled with experiment in a many-body approach\cite{Allen,Held},
showing that even if the $4f$ level is far below $E_F$ it can form a Kondo-like
many-body resonance at $E_F$ leading to very large values of the susceptibility 
and the specific heat. This illustrates how an oversimplified (mean-field) method 
can nevertheless lead to reasonable results if it is combined with a bad choice 
of parameters. Correcting the parameters then forces  us to use a better method 
and to find out more about the correct physics.

\section{Projecting out states}\label{sec:2}
\subsection{One-particle Hamiltonian}\label{sec:2.1}

One approach to the construction of models is to project out states\index{project out states} which are
believed to to not be essential for the physics. We can illustrate this
for a one-particle Hamiltonian
\begin{equation}\label{eq:2.1}
H=\sum_i \varepsilon_in_i +\sum_{i \ne j}t_{ij} \psi^{\dagger}_i\psi_j
\end{equation}
We introduce a projection operator
\begin{equation}\label{eq:2.2}
P=\sum_{\nu}|\nu\rangle\langle \nu|,
\end{equation}
where $|\nu\rangle$ are states we want to keep. We introduce        
the resolvent operator
\begin{equation}\label{eq:2.3}
(z-H)^{-1}=\sum_{\nu}|\nu\rangle\langle\nu|(z-H)^{-1}
\sum_{\mu}|\mu\rangle \langle \mu|=
\sum_{\nu}|\nu\rangle {1\over z-E_{\nu}}
\langle \nu|,
\end{equation}
which has poles for $z=E_{\nu}$ at the eigenvalues.
Introducing the complement $Q=(1-P)$, we can write the Hamiltonian as\cite{Lowdin,downfolding}
\begin{equation}\label{eq:2.3a}
\left(\begin{array}{cc}
 H_{PP}  & H_{PQ} \\
 H_{QP} & H_{QQ} \\
\end{array}\right),
\end{equation}
where, e.g., $H_{PP}=PHP$.
Then we can derive the exact result 
\begin{equation}\label{eq:2.4}
P(z-H)^{-1}P=\lbrack z-H_{PP}-H_{PQ}(z-H_{QQ})^{-1}H_{QP}\rbrack^{-1}.
\end{equation}
The operator $P(z-H)^{-1}P$ has the same poles as the original operator $(z-H)^{-1}$,
if the corresponding eigenstates have weight inside the space $P$. The new operator
has a smaller dimension, but because of the $z$ dependence it is not simpler.
To simplify the expression,  we put $z$ equal to an energy ($\varepsilon_0$) in
the range of interest. The operator is then energy independent. As an additional 
simplification, we may assume that the off-diagonal elements of $H_{QQ}$ can be neglected. 
Then the matrix elements of the new operator become
\begin{equation}\label{eq:2.5}
t_{ij} \to t_{ij}-\sum_{\mu \in Q} { t_{i\mu}t_{\mu j} \over \varepsilon_0-E_{\mu}}.
\end{equation}
This latter approximation is accurate if the states being projected out are
much higher in energy that the states of interest and if the off-diagonal elements are
small compared with the energy difference $\varepsilon_0-E_{\mu}$. The assumption
about $H_{QQ}$ being diagonal can also be relaxed. This approach reduces the size of
the Hamiltonian matrix (reduces the number of states) at the cost of obtaining 
more long-range hopping. For a one-particle Hamiltonian, this approach is a controlled
and systematic procedure for reducing the size of the Hamiltonian.

\subsection{Many-body Hamiltonian}\label{sec:2.2}

We now consider a many-body Hamiltonian, with a two-body interaction in 
the form of a Coulomb interaction. We then define $P$ as projecting out
states\index{project out states} that have no electron in certain (high-lying) one-particle states 
$|\mu\rangle$ and $Q=1-P$. We consider a Coulomb interaction with four 
(creation and annihilation) operators and project out a state with one 
electron in $|\mu\rangle$. Then $H_{QP}$ contains an operator 
$c^{\dagger}_{\mu\sigma}$ an $H_{PQ}$ and operator $c_{\mu\sigma}$.
Even if we assume $H_{QQ}$ to be diagonal, we are left with an operator
$H_{PQ}H_{QP}$ acting on a state without electrons in $|\mu\rangle$. Then 
$c_{\mu\sigma}c^{\dagger}_{\mu\sigma}\equiv 1$, and two operators drop out.
But we are still left with six other operators, which in the general case 
are all different. We have then generated a three-body operator. This is 
too complicated, and all such operators need to be neglected. Unless it
can be shown that these terms are small,
this means that there is not a controlled systematic procedure for reducing 
the number of states. We then have to rely on more intuitive approaches.

\begin{figure}[h]
 \centering
 \includegraphics[width=0.4\textwidth]{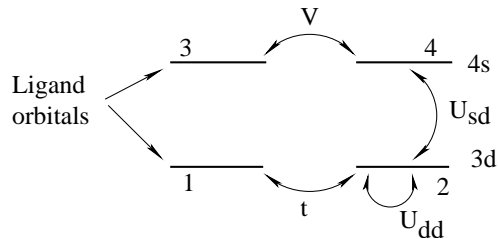}
 \caption{\label{fig:3d}Schematic picture of a very simple model of a 
transition metal compound, with a $3d$ atom (levels 2 and 4) coupling 
to a ligand (with levels 1 and 3).
}
\end{figure}

As a simple example we consider a very simple model which is relevant for $3d$ impurities.
The model is constructed so that an exact solution can be found. We want to illustrate
how this model can be projected down to a simpler model with renormalized parameters. 
We introduce the Hamiltonian\cite{parameters}
\begin{equation}\label{eq:3.1}
H=\sum_{\sigma}\lbrack \sum_{i=1}^4 \varepsilon_in_{i\sigma}
+(t\psi_{1\sigma}^{\dagger}\psi^{\phantom \dagger}_{2\sigma}+V\psi_{3\sigma}^{\dagger}
\psi^{\phantom \dagger}_{4\sigma}+{\rm H.c.})\rbrack\\+U_{dd}n_{2\uparrow}n_{2\downarrow}+
U_{sd}\sum_{\sigma\sigma^{'}}n_{2\sigma^{\phantom '}}n_{4\sigma^{'}} 
\end{equation}
where level 2 corresponds to a $3d$ level and level 4 to a $4s$ level
on a transition metal atom. Level 1 and 3 correspond to a ligand 
coupling to the $3d$ atom via the hopping integrals $t$ and $V$.
On the $3d$ atom there is a large Coulomb interaction $U_{dd}$ between
electrons in the $3d$ level and a weaker $U_{sd}$ interaction between 
the $3d$ and $4s$ levels. We assume that orbital 2 is quite localized,
so that $t$ is small, but that levels 3 and 4 are delocalized, so that 
$V$ is large. The level structure is shown schematically in Fig.~\ref{fig:3d}. 

We first consider the spinless case, and put one electron in each
of the spaces 1+2 and 3+4. We derive parameters in an intuitive
approach, and then compare with a controlled projection approach, possible
in this case. We introduce the eigenstates of the space 3+4 with the     
electron in space 1+2 on site 1 or 2. With the electron on level 1            
the bonding and antibonding eigenstates are  
\begin{eqnarray}\label{eq:3.2}
&&\psi_{b1}=a_3\psi_3+a_4\psi_4 \\ 
&&\psi_{a1}=a_4\psi_3-a_3\psi_4,\nonumber
\end{eqnarray}
with the energies $\varepsilon_{b1}$ and $\varepsilon_{a1}$.
With the electron in 1+2 on level 2 the states are
\begin{eqnarray}\label{eq:3.3}
&&\psi_{b2}=({\rm cos} \phi) \psi_{b1}+({\rm sin} \phi)\psi_{a1} \\
&&\psi_{a2}=({\rm sin} \phi) \psi_{b1}-({\rm cos} \phi)\psi_{a1},\\
\end{eqnarray}
with the energies $\varepsilon_{b2}$ and $\varepsilon_{a2}$.
Here $\phi$ is of the order $U_{sd}/V$ which we is small in the limit we consider below.
We assume that the electron in the space 3+4 can adjust completely to the 
movement of the electron in space 1+2 due to $|V|>>|t|$. We then replace the
four-level model in Eq.~(\ref{eq:3.1}) by a two-level model with the 
effective level positions
\begin{equation}\label{eq:3.3a}
\varepsilon_1^{\rm eff}=\varepsilon_1+\varepsilon_{b1}; \hskip 0.5cm \varepsilon_2^{\rm eff}=\varepsilon_2+\varepsilon_{b2}
\end{equation}

To test this, we now solve the full model exactly. We introduce a complete basis set
\begin{eqnarray}\label{eq:3.4}
&&|\tilde 1\rangle=\psi^{\dagger}_1\psi^{\dagger}_{b1}|0\rangle \nonumber \\
&&|\tilde 2\rangle=\psi^{\dagger}_2\psi^{\dagger}_{b2}|0\rangle \\
&&|\tilde 3\rangle=\psi^{\dagger}_1\psi^{\dagger}_{a1}|0\rangle \nonumber \\
&&|\tilde 4\rangle=\psi^{\dagger}_2\psi^{\dagger}_{a2}|0\rangle,\nonumber
\end{eqnarray}
where we have chosen the basis set so that only the first two states are relevant 
if the assumptions above are correct. We now calculate the resolvent operator \cite{parameters}
\begin{equation}\label{eq:3.5}
(z-H)^{-1}=\left(\begin{array}{cccc}
z-\varepsilon_1-\varepsilon_{b1} & -t \ {\rm cos} \phi & 0 & t \ {\rm sin} \phi \\ 
-t \ {\rm cos} \phi & z-\varepsilon_2-\varepsilon_{b2} & - t \ {\rm sin} \phi & 0 \\
0 & - t \ {\rm sin} \phi & z-\varepsilon_1-\varepsilon_{a1} & -t \ {\rm cos} \phi  \\
t \ {\rm sin} \phi & 0 & -t \ {\rm cos} \phi & z-\varepsilon_2-\varepsilon_{a2}  \\
\end{array}\right)^{-1}.
\end{equation}
We now focus on the upper left $2\times 2$ corner and use L\"owdin folding \cite{Lowdin}
to project out the two high-lying states\index{project out states}. For instance, the 11 element takes the form
\begin{equation}\label{eq:3.6}
\tilde H_{11}=\varepsilon_1+\varepsilon_{b1}+{t^2 (z-\varepsilon_1-\varepsilon_{a1}){\rm sin}^2 \phi \over 
(z-\varepsilon_1-\varepsilon_{a1})(z-\varepsilon_2-\varepsilon_{a2})-t^2{\rm cos}^2 \phi}.
\end{equation}
For simplicity, we put $\varepsilon_1=\varepsilon_2$ and assume that the term $t^2{\rm cos}^2 \phi$
in the denominator can be neglected. Putting $z\approx \varepsilon_1+\varepsilon_{b1}$, we then find
that the correction term in Eq.~(\ref{eq:3.6}) is of the order $t(t/V)(U_{sd}/V)^2$. If $|V| \gg |t|$
and $|V| \gg U_{sd}$, it is indeed justified to neglect the correction term.
We then find that the level positions difference, $\varepsilon_1^{\rm eff}-\varepsilon_2^{\rm eff}$,
have corrections to zeroth order in $(1/V)$, 
due to $\varepsilon_{b1}$ and $\varepsilon_{b2}$. These corrections are included 
in our intuitive approach above. Then there is a second order correction to the hopping
integral\index{hopping integral}  due to ${\rm cos}\phi$. This correction is due to the fact that the electron 
in the space 3+4 cannot completely follow the electron in space 1+2 in the optimum way. This correction
is usually neglected. 

We now turn to the same model with spin degeneracy. The exact solution can then be obtained from
a $16\times 16$ matrix. In this case the analytical calculation is to complicated to 
illustrate what happens, and we focus on a numerical calculation. We first calculate the energy $E(n_2)$
of the 3+4 space as a function of the occupancy of level 2. We then obtain 
\begin{eqnarray}\label{eq:3.6a}
&&\varepsilon_1^{\rm eff}=\varepsilon_1+E(0) \nonumber\\
&&\varepsilon_2^{\rm eff}=\varepsilon_2+E(1)\\
&&U^{\rm eff}=E(2)+E(0)-2E(1) \nonumber
\end{eqnarray}
in analogy with the spinless case. We then calculate the ground-state energy, $E_0$, the occupancy of 
level 2, $n_2$ and the spin susceptibility $\chi=-\partial^2E_0(H)/\partial H^2$, where the 
model couples to an external magnetic field via the term $-H(n_{2\uparrow}-n_{2\downarrow}$. 
The results are shown in Table~\ref{table:fourlevel}. We have added a contribution $2V$
to the total energy $E_0$, since there would have been a trivial contribution $-2V$ if there had
been no interaction between spaces 1+2 and 3+4. As expected, the agreement between the approximate
(Renorm.) and exact results improve as $|V|$ is increased. However, the agreement is surprisingly good even for
$V=t$.
\begin{table}
\caption{\label{table:fourlevel}Ground-state energy $E_0$, occupancy of level 2, $n_2$,  and susceptibility
$\chi$ of the spin-degenerate model (\ref{eq:3.1}). The parameters are $\varepsilon_1=\varepsilon_2=
\varepsilon_3=\varepsilon_4=0$, $t=1$, $U_{dd}=4$ and $U_{sd}=2$.}
\begin{tabular}{ccccccccc}
\hline
\hline
$V$ & $\varepsilon_2^{\rm eff}$-$\varepsilon_1^{\rm eff}$ & $U^{\rm eff}$ & \multicolumn{2}{c}{$E_0+2V$} & \multicolumn{2}{c}{$n_2$} & \multicolumn{2}{c}{$\chi$} \\
  &   &   & Renorm. & Exact & Renorm. & Exact & Renorm. & Exact \\
\hline
1.0 &  1.17 & 3.18 & -1.05 & -0.95 & 0.380 & 0.364 & 0.314 & 0.312 \\
1.5 &  1.39 & 3.21 & -0.97 & -0.90 & 0.339 & 0.326 & 0.266 & 0.262 \\
2.0 &  1.53 & 3.29 & -0.92 & -0.88 & 0.317 & 0.307 & 0.240 & 0.237 \\
3.0 &  1.68 & 3.44 & -0.87 & -0.85 & 0.292 & 0.287 & 0.214 & 0.213 \\
4.0 &  1.75 & 3.55 & -0.85 & -0.84 & 0.280 & 0.277 & 0.202 & 0.201 \\
6.0 &  1.83 & 3.68 & -0.83 & -0.82 & 0.268 & 0.267 & 0.190 & 0.190 \\
10.0 &  1.90 & 3.80 & -0.81 & -0.81 & 0.259 & 0.258 & 0.181 & 0.181 \\
20.0 &  1.95 & 3.90 & -0.80 & -0.80 & 0.252 & 0.252 & 0.174 & 0.174 \\
\hline
\hline
\end{tabular}
\end{table}

\section{Effective Coulomb interaction}\label{sec:4}

The essential point in the model in the previous section is that we 
can distinguish between two types of electrons, ``slow'' electrons (space 1+2)
''fast'' electrons (space 3+4), in the following referred to as ``localized'' 
and ``delocalized''.  The idea is that the delocalized electrons 
are assumed to adjust in an optimum way to the movements of the localized 
electrons. We can then estimate effective parameters in a similar way 
as in the previous section. For each system we then have to decide 
which electrons are localized and included explicitly in the model and
which are delocalized and only included implicitly as a renormalization of the parameters\index{renormalization of the parameters}.
This is illustrated in Table~\ref{table:4.1}. For $4f$ compounds the $4f$ DFT band width 
is about 1/10 of the $5d$ band width, and we may reasonably talk about two types
of electrons. For $3d$ compounds this distinction is much less clear cut.

\begin{table}
\caption{\label{table:4.1}''Slow'' (``localized'') and ''fast'' (``delocalized'') 
electrons for $3d$ and $4f$ compounds}
\begin{tabular}{ccc}
\hline
\hline
System   &  Localized & Delocalized \\
\hline
$4f$ compounds & $4f$   & $5d$  \\
$3d$ compounds & $3d$  & $4s$, $4p$ \\
\hline
\hline
\end{tabular}
\end{table}

\subsection{``Perfect screening''}\label{sec:4.1}

We now focus on the calculation of an effective Coulomb integral $U^{\rm eff}$\index{Coulomb integral},
as an essential model parameter\index{model parameter}.
We apply the approach in the previous section to real system. For that reason,
we need to know how the energy of the system varies  with the occupancy 
of, e.g., a $3d$ or $4f$ level [Eq.~(\ref{eq:3.6a})].
Herring\cite{Herring} estimated these energies using atom data, assuming that any change
in the number of localized electrons on an atom is compensated by the opposite change in 
the number of delocalized electrons on the same atom.
For a $3d$ metal this can be written as
\begin{equation}\label{eq:4.0}
U=E(3d^{n+1}4s^0)+E(3d^{n-1}4s^2)-2E(3d^n4s^1),
\end{equation}
where $E(3d^n4s^m)$ is the energy of an atom (ion) with $n$ 3d electrons and 
$m$ 4s electrons. In this approach is is assumed that the variation in the
number of $3d$ electrons is perfectly screened by a change in the number of $4s$ electrons.
We refer to this as ``perfect screening''\index{screening}.

A similar method was used by Cox {\it et al.} \cite{Cox} who studied transition metals 
and Herbst {\it et al.}\cite{Herbst} who studied the rare earths. They performed Hartree-Fock 
calculations for renormalized atoms with Wigner-Seitz boundary conditions.

\subsection{Constrained density functional formalism}

Dederichs {\it et al.}\cite{Dederichs} calculated $U$ using a constrained
density functional formalism\index{constrained density 
functional formalism}. The functional for a $3d$ compound is written as
\begin{equation}\label{eq:4.1}
E\lbrack n_{3d}^i\rbrack = F\lbrack n \rbrack + \int d^3r 
V_{ext}({\bf r})n({\bf r})+\mu\lbrace \int d^3r n({\bf r})-N\rbrace
+\mu_{3d}^i \lbrace \int d^3r n_{3d}^i({\bf r})-n_{3d}^i\rbrace.
\end{equation}
Here $F[n]$ describes the kinetic and potential energy of the system, $V_{ext}({\bf r})$
is an external potential, $\mu$ is the chemical potential and $\mu_{3d}^i$ is 
Lagrange parameter. $n_{3d}^i$ is the number of localized electrons on site $i$,
referred to as the central site in the following. A stationary point of the energy 
functional to density variations is searched
\begin{equation}\label{eq:4.2}
0={\partial F\over \partial n}+V_{ext}({\bf r})+\mu+ \mu_{3d}^iP_{3d}^i,
\end{equation}
where $P_{3d}^i$ is a projection operator acting on the localized electrons.
From this a new Kohn-Sham equation can be derived, where $\mu_{3d}^i$
enters as an additional nonlocal potential acting only on the localized electrons. 
$\mu_{3d}^i$ is varied until the prescribed number of $3d$ electrons is obtained.
This requires
a definition of localized  electrons. For instance, in methods based on an expansion
in spherical waves in a region around each nucleus a natural definition can be introduced. 
This way an effective $U^{\rm eff}$ is calculated, using a formula equivalent to Eq.~(\ref{eq:3.6a}).

In the approach above, $U^{\rm eff}$ contains a change of the kinetic energy of the electrons
included explicitly in the model. This contribution needs to be subtracted. Hybertson 
{\it et al.}\cite{Schluter} did this by considering the model Hamiltonian in which
 $U^{\rm eff}$ will be used, e.g.,                      
\begin{equation}\label{eq:4.3}
H=\sum_{ij\sigma}t_{ij}\psi^{\dagger}_{i\sigma}
\psi^{\dagger}_{j\sigma}+\sum_i U^{\rm eff}n_{i\uparrow}n_{i\downarrow},             
\end{equation} 
where $t_{ij}$ are hopping integrals. This model is then solved in a constrained mean-field 
theory, to simulate the constrained density functional approach. The energy as a function 
of the constrained occupancies is calculated, and $U^{\rm eff}$ is varied until the 
constrained DFT result is reproduced. We refer to this as cLDA.  Cococcioni and 
Giroconcoli\cite{Cococcioni} used a similar approach.

An alternative approach was used by McMahan {\it et al.}\cite{McMahan} and by Gunnarsson 
{\it et al.}\cite{cut}.  They performed a band structure calculation with a 
large unit cell\cite{McMahan,cut,Anisimov}. Then the hopping integrals from the
orbital with localized electrons is cut off for the central atom in the unit cell. 
Then the occupation of the orbital can be trivially varied without a variation 
of the kinetic energy for hopping in and out of the orbital, since this energy is 
zero. Double-counting is also explicitly avoided, contrary to claims elsewhere\cite{note}. 
This method is referred to as ``cut off' LDA. In a different method, the hopping 
between the localized orbitals and the delocalized orbitals was cut on all sites,
not only on the central site\cite{Imada}.

\subsection{Constrained RPA}\label{sec:4.3}

A different approach was taken by Aryasetiawan {\it et al.}\cite{Ferdi}.
They calculated the Coulomb interaction using a constrained random phase
(RPA) screening\index{screening}. In RPA the polarizability is written as
\begin{eqnarray}\label{eq:4.4}
&&P({\bf r},{\bf r}^{'}:\omega)=\sum_i^{\rm occ}\sum_j^{\rm unocc}
\psi_i({\bf r})\psi_i^{\ast}({\bf r}^{'})\psi_j^{\ast}({\bf r})\psi_j({\bf r}^{'}) \\ 
&&\times ({1\over \omega-\varepsilon_j+\varepsilon_i+i0^{+}}-{1\over \omega+\varepsilon_j-\varepsilon_i-i0^{+}}
),\nonumber
\end{eqnarray}
where $\psi_i({\bf r})$ and $\varepsilon_i$ are one-particle eigenfunctions and eigenvalues.
Calculating the effective Coulomb interaction by using this screening would be 
incorrect, since it would involve double-counting. The Hubbard model explicitly allows 
localized electrons to screen the interaction between localized electrons, and the use 
of Eq.~(\ref{eq:4.4}) would then lead to double counting. Aryasetiawan {\it et al.}\cite{Ferdi} 
therefore excluded contributions to Eq.~(\ref{eq:4.4}) where both $i$ and $j$ stand 
for Bloch states containing mainly localized states.
For a transition metal compound they then excluded states within an energy window where the states
are mainly of $3d$ character and for a rare earth compound a window where the states are mainly of 
$4f$ character. The definition of the energy window involves  uncertainties \cite{Ferdi}.
This method is  referred to as cRPA. 

\subsection{Screening and breathing}\label{sec:4.4}

The definition of $U$ can be approximately rewritten
as 
\begin{equation}\label{eq:4.5}
U=E(n+1)+E(n-1)-2E(n)\approx {\delta \varepsilon \over \delta n},
\end{equation}
where $E(n)$ is the energy of the system with $n$ localized electrons and 
$\varepsilon$ is the energy eigenvalue of the localized orbital and 
$n$ is the occupancy. If the system were not allowed to relax, this would
lead to $U=F$, where $F$ [Eq.~(\ref{eq:1.2})] is the direct Coulomb integral of the orbital. 
In reality, the charge density relaxes and the corresponding change in the 
electrostatic potential acts back on the orbital eigenvalue, reducing the shift
as $n$ is varied and leading to a renormalized $U$. We can illustrate 
this for the case of a Mn impurity in CdTe \cite{Mn}. First a free Mn atom
is studied [Table~\ref{table:4.1a}]. The spherical part 
$F^0_{3d,3d}$ of the direct Coulomb integral is large, 21 eV. The main
renormalizing process is a breathing of the $3d$ orbital, where the orbital
expands as the $3d$ occupancy is increased \cite{Mn,Solovyev}. This reduces $U$ by about 5 eV.
Breathing of the $4s$, $4p$ and core orbitals contribute less. The net result
is a reduction of $U$ from about 21 eV to about 13 eV. In the solid there are 
similar breathing effects, reducing $U$ to about 15 eV 
(see Table~\ref{table:4.2}). However, now there is additional charge transfer 
from the surrounding to the Mn atom, reducing the $U$ by almost 8 eV. Charge 
transfer to near neighbors (n.n.) plays a smaller role. The result is reduction 
of $U$ to about 7 eV according to this calculation.
\begin{table}
\caption{\label{table:4.1a}Contribution to $U$ for a free Mn atom
with the configuration $3d^{5.1}4s^{0.64}4p^{0.70}$. This corresponds to 
the configuration for Mn in CdTe.
}
\begin{tabular}{ll}
\hline
\hline
Unrenormalized ($F^0$) &  21.4 eV \\
Relaxation of $3d$ orbital & -5.2 eV \\
Relaxation of $4s$, $4p$ orbitals & -2.2 eV \\
Relaxation core, XC effects & -1.2 eV \\
\hline
Atomic $U$ & 12.8 eV \\
\hline
\hline
\end{tabular}
\end{table}
\begin{table}
\caption{\label{table:4.2}Contribution to $U$ for a Mn impurity in CdTe.
}
\begin{tabular}{ll}
\hline
\hline
On-site relaxation & 15.4 eV \\
Charge transfer from Mn & -7.6 eV \\
Charge transfer to n.n. ligand & -0.4 eV \\
\hline
Solid state $U$  & 7.4 eV \\
\hline
\hline
\end{tabular}
\end{table}

The breathing effect can also be understood from Slater's rules\cite{Slater}.
According to these rules, the effective nuclear charge for a $3d$ orbital is 
$Z^{\ast}=Z-18-0.35(n_{3d}-1)$, where $Z$ is the true nuclear charge and $n_{3d}$ 
is the $3d$ occupancy. This illustrates how the effective nuclear charge is reduced and 
the orbital expands as $n_{3d}$ is increased. According to Slater's rules,
the occupancy of the $4s$ and $4p$ orbitals do not influence $Z^{\ast}$ for the
$3d$ orbital. This then suggest that the charge transfer in the solid to $4s$ and
$4p$ should not influence breathing very much. This is also supported by
a comparison of Tables~\ref{table:4.1a} and \ref{table:4.2}.

\subsection{Results}\label{sec:4.5}

We now consider results obtained using the methods above for $3d$ and $4f$
metals. Specifically, we consider Fe and Ce as examples of $3d$ and $4f$ metals.
The results are shown in Table~\ref{table:4.3}. ``Perfect screening'' provides a rather 
good estimate for both Fe and Ce. The ''cut-off'' method gives a substantially too large 
$U$ for Fe. It was found\cite{Anisimov} that only about half the screening\index{screening} 
charge is on the Fe atom, as one would expect from the energetics of the screening
process\cite{Anisimov}. It is then to be expected that $U$ is substantially larger than 
the ``perfect screening'' result. cLDA gives a very good result compared with
experiment, and actually somewhat smaller than ``perfect screening''. It is
not clear why this result is so different from the ''cut-off'' method, and it would be
interesting to study the screening in cLDA. The $U$ in cRPA is a bit too large. 
Interestingly, the ''cut-off'' method gives a good estimate of $U$ towards the end
of the $3d$ series, e.g. for the cuprates\cite{cuprates}.

For Ce ``perfect screening'' provides a fairly accurate estimate of $U$. The ``cut off'' 
method gives only a slightly larger $U$, in good agreement with experiment. In this
case it is found that the screening charge on Ce is approximately unity\cite{Anisimov}, 
so it is not surprising that there is rather good agreement with ``perfect screening''.
cLDA gives a $U$ slightly smaller than ``perfect screening'' and $U$ in cRPA is 
substantially smaller. It is not clear why cRPA implies such an effective screening  
and gives a $U$ that is only roughly half the experimental estimate.

\begin{table}
\caption{\label{table:4.3}Results for $U$ for Fe as an example of a $3d$ metal       
and Ce illustrating  a $4f$ metal.
}
\begin{tabular}{cccccc}
\hline
\hline
System   & cLDA & ``cut-off'' & cRPA &``perfect screening'' & Exp \\
\hline
Fe & 2.2\cite{Cococcioni} &   6\cite{Anisimov}      &  4\cite{Ferdi} & 2.7\cite{Cox} & 2\cite{Sasha,Drchal} \\
Ce     &  4.5\cite{Cococcioni}  &   6\cite{Anisimov}    &  3.2-3.3\cite{Ferdi} & 5\cite{Herbst}  &  5-7\cite{AdvPhys} \\
\hline
\hline
\end{tabular}
\end{table}

\section{Neglected renormalizations}\label{sec:5}

In this section we discuss two renormalizations of parameters\index{renormalization of parameters}, 
which are usually neglected. 
The purpose is not argue that these effects should be included. This could be done, 
but it would result in more parameters and the results would probably be less transparent. 
The purpose is rather to illustrate that the parameters of effective models contain
complicated renormalizations, and that {\it ab initio} estimates of such parameters
may neglect several such effects. The purpose is also to show that if we insist
on rather simple model, which is advocated here, the effective parameter may
actually be different for different properties. 

\subsection{Configuration dependence of hopping integrals}\label{sec:5.1}

We already discussed in Sec.~\ref{sec:4.4} that there is a substantial
breathing of the localized orbital when the occupancy is changed. 
This changes the hopping integral\index{hopping integral} into this orbital. 
In the LMTO method \cite{LMTO}, used here, the hopping integral $V$ is related 
to a potential parameter $\tilde \Delta$,
\begin{equation}\label{eq:5.1}
V^2 \sim \tilde \Delta\approx {s\over 2} [\phi_l(C,s)]^2,
\end{equation}
where $\phi_l(C,s)$ is the value of the localized orbital at the Wigner-Seitz radius $s$.
The localized orbital with the angular momentum $l$ is solved for an energy $C$, which
gives the logarithmic derivative $-l-1$ at $s$. The value of $\tilde \Delta$ is shown 
in Table~\ref{table:5.1} for a few metals with and without a core hole \cite{config}. 
The table illustrates the strong dependence of the hopping on the configuration used to 
calculate $\tilde \Delta$. For instance, if we want to describe how a host electrons hop 
into a Ce atom, should we then use the initial configuration or  the final configuration 
to calculate $\tilde \Delta$ or should we use an average? Table~\ref{table:5.1} shows that 
the difference could even be as much as a factor of two.

\begin{table} 
\caption{\label{table:5.1}Potential parameter $\tilde \Delta$ for different configurations
of Mn, Ce and U in non-spin-polarized calculations. The localized orbital is $3d$ (Mn), $4f$ (Ce)
and $5f$ (U), and we consider a core hole in the $1s$ (Mn), $3d$ (Ce) and $4f$ (U) orbital.
The occupancy of the localized and core orbital is $n_l$ and $n_c$, respectively.
We introduce $n_l^0$, which is 5 (Mn), 1 (Ce) and 3 (U) and $n_c^0$ which is 2 (mn), 
10 (Ce) and 14 (U). All energies are in Ry.
}
\begin{tabular}{ccccc}
\hline
\hline
$n_l$ & $n_c$ & Mn & Ce & U \\
\hline
$n_l^0-1$ & $n_c^0$ & 0.0051 & 0.0008 & 0.0072 \\
$n_l^0$ & $n_c^0$ & 0.0085 & 0.0019 & 0.0091 \\
$n_l^0+1$ & $n_c^0$ & 0.0129 & 0.0038 & 0.0112 \\
$n_l^0$ & $n_c^0-1$ & 0.0040 & 0.0005 & 0.0053 \\
$n_l^0+1$ & $n_c^-10$ & 0.0067 & 0.0011 & 0.0069 \\
\hline
\hline
\end{tabular}
\end{table}

To address this issue, we temporary introduce an impurity model with two orbitals\cite{config}
\begin{eqnarray}\label{eq:5.2}
&&\phi_l^0\equiv \phi_l(r,n_l^0) \\
&&\phi_l^1\equiv A {\partial \over \partial n_l}\phi_l(r,n_l)|_{n_l=n_l^0},\nonumber
\end{eqnarray}
where $A$ is chosen so that $\phi_l^1$ is normalized. By forming linear combinations of
$\phi_l^0$ and $\phi_l^1$, we can obtain an appropriate orbital for different occupancies,
i.e., describing breathing. In, for instance, an Anderson impurity model we then introduce
a term leading to transitions between these two orbitals
\begin{equation}\label{eq:5.3}
\tilde U \sum_{m\sigma}(\psi^{\dagger}_{1m\sigma}\psi^{\phantom \dagger}_{0m\sigma}+{\rm H.c.})(n_0+n_1-n_l^0),
\end{equation}
where $n_i=\sum_{m\sigma}n_{im\sigma}$. If the occupancy of the two levels adds up
to $n_l^0$, the orbital $\phi_l^0$ is appropriate and there is no mixing of the orbital $\phi_l^1$.
For any other occupancy transitions to $\phi_l^1$ are induced and the system has
the freedom to adjust to the occupancy. For Mn in CdTe we find that $\tilde U=0.16$ Ry.
The energies of the two orbitals are quite different, $\varepsilon_0=-0.45$ Ry and 
$\varepsilon_1= 1.68$ Ry. The model then tends to have two sets of states, one set at
$\varepsilon_0$ and one set at $\varepsilon_1$. We can then project out all high-lying
states, having a substantial weight in $\phi^1_l$. The result is then that we recover the 
normal Anderson impurity model, with just one localized orbital. But in this process 
the hopping matrix elements are modified. Since the mixing matrix element 
$\tilde U/(\varepsilon_1-\varepsilon_0)=0.08\ll1$,   this approach should be rather accurate.

We can then answer the question of how to 
calculate these elements. Let us consider a host electron hopping into a configuration 
with $n_l$ localized electrons, resulting in a configuration with $n_l+1$
localized electrons. The projection procedure then shows that the matrix element should be 
calculated using $n_l+1$ electrons, i.e., the end configuration\cite{config}. For $n_l=0$ this 
is easy to understand. In the initial state there is no localized electron and the 
extent of the localized wave function then plays no role. It is then natural
that it is the wave function in the final configuration that matters. In a similar 
way it is the initial configuration that matters when an electron hops out of the 
localized orbital.

We then should be using different hopping integrals for different experiment. 
For Ce compounds, for instance, $f^0-f^1$-hopping is particularly important
for valence photoemission, and we would use $n_l=1$ for calculating these 
hopping matrix elements. For inverse photoemission, we are often interested in
the relative weights of the $f^1$ and $f^2$ peaks. We then need to distinguish
between the calculation of the ground state and the calculation of the final
states, resulting from the inverse photoemission process. In the ground-state 
calculation the important matrix elements would be calculated for $n_l=1$ and 
in the final state for $n_l=2$.  For core level 
spectroscopies we should in addition include a core hole for the calculation of
matrix elements for the final states but not for the ground-state.

As argued above, this would lead to a complicated model. It seems questionable if
the possible additional gain in physics would justify such a complicated model
with additional parameters. However, the example illustrates one source of
uncertainty in models with configuration independent hopping parameters. 
It also illustrates how parameters can be different for different experiments.

\subsection{Many-body renormalization of hopping integrals}\label{sec:5.2}

In Secs.~\ref{sec:2.2} and \ref{sec:4} we discussed how the effective level
energies and Coulomb integrals can be obtained by letting delocalized 
electrons adjust to the movements of localized electrons. This approach,
however, raises questions about other many-body effects. One issue is the
Anderson orthogonality catastrophe \cite{Andersonorthog}. Consider the case when 
delocalized electrons interact with a (truly) localized electron via
a Coulomb interaction. Let us then change the occupancy of the localized
level by one and let $|0\rangle$ and $|1\rangle$ be the lowest states 
of the delocalized electrons in the presence of 0 or 1 localized
electrons, respectively. The $\langle0|1\rangle=0$ for an infinite system
\cite{Andersonorthog}. One might then think that the hopping integrals\index{hopping integrals} should
be reduced by such effects. When a delocalized electron hops into a localized
level, all the other electrons would adjust their wave functions to the 
new potential. Then one might expect that the overlap $\langle0|1\rangle=0$ enters 
the effective hopping integral. This is, however, not the appropriate comparison.
Anderson's orthogonality catastrophe\index{Anderson orthogonality catastrophe} 
refers to the case when the localized
electron is removed from the system. Here it hops into or out of delocalized 
states. The appropriate comparison is then X-ray absorption (XAS) or X-ray 
emission (XES).  In addition to the Anderson effect there is then an exciton 
like effect, transferring spectral weight towards the Fermi energy. For instance, 
the XES spectrum looks like
\begin{equation}\label{eq:5.3a} 
S(\omega)\sim ( {\tilde \omega \over \omega-\omega_0})^{\alpha}
\Theta(\omega-\omega_0),
\end{equation}
where $\tilde \omega$ is a typical energy and $\omega_0$ is the threshold energy.
The exponent $\alpha$ is positive and determined by the phase shift due
to the Coulomb interaction between localized and delocalized electrons.
From Eq.~(\ref{eq:5.3a}) we might then expect hopping integrals for states 
close to the Fermi energy to be enhanced. This would then in particular 
influence thermodynamic properties.

To check these ideas we have considered the spinless model\cite{xas}
\begin{equation}\label{eq:5.4}
H=\sum_{k=1}^N\varepsilon_kn_k+\varepsilon_dn_d+{t\over \sqrt{N}}\sum_{k=1}^N
(\psi_k^{\dagger}\psi^{\phantom \dagger}_d+{\rm H.c.})+{U_{sd}\over N}\sum_{k=1}^N\sum_{l=1}^N \psi^{\dagger}_k\psi^{\phantom \dagger}_ln_d,
\end{equation}
where we have introduced $N$ delocalized states with the energies $\varepsilon_k$ and a localized state 
with the energy $\varepsilon_d$. There is a hopping integral $t$ connecting the localized and delocalized 
states. When the localized level is occupied the delocalized electrons feel a scattering
potential $U_{sd}$. The delocalized levels are equally spaced over an energy $2B$.

This model can be solved using exact diagonalization for finite $N$ \cite{xas}. We 
have used $N=17$ and the number of electrons $N_{el}=9$. Although this is far from 
an infinite system, Anderson's orthogonality catastrophe already has an effect. For 
$B=5$, $\varepsilon_d=-1.5$ and $U_{sd}=5$, the overlap between the lowest states of 
delocalized electrons in the presence or absence of a localized electron is $0.85<1$.  
We then calculate the energy lowering $\Delta E=E_0-\sum_k^{'} \varepsilon_k$, where 
$E_0$ is the ground-state energy and the sum goes over the $N_{el}$ lowest states. 
We also calculate the $3d$ occupancy, $n_{d}$  and the charge susceptibility 
$\chi_c=-\partial n_d/\partial \varepsilon_d$. The results are shown in 
Table~\ref{table:5.2} and \ref{table:5.3}. 

\begin{table}
\caption{\label{table:5.2}Energy lowering $\Delta E$ and occupancy of the $d$ level $n_d$
in the exact calculation (``Ex.'') compared with results of calculations for the model
(\ref{eq:5.4}) with $U_{sd}=0$. The unrenormalized $d$ level position was used for ``Unre.'' 
and the calculated renormalized position for ``Ren.'' and ``XAS''. For ``XAS'' the effective 
hopping integral was renormalized [Eq.~(\ref{eq:5.5})] and for ``Fit'' both the level position
and the hopping were adjusted to obtain the best agreement with the exact results.
The parameters are $t=1$, $B=5$ $N=17$ and $N_{\rm el}=9$. 
}
\begin{tabular}{ccccccccccccccc}
\hline
\hline
 &  & \multicolumn{5}{c}{$-\Delta E$} & \multicolumn{5}{c}{$n_d$} &  & &  \\
$\varepsilon_d$    & $U_{sd}$  & Ex. & Ren. & Unre. & Fit & XAS & Ex. & Ren. & Unre. & Fit & XAS & $\varepsilon_d^{\rm calc}$ & $\varepsilon_d^{\rm fit}$ & $t_{\rm eff}^{\rm fit}$\\
\hline
-1.5 & 1 & 1.33 & 1.28 & 1.66 & 1.33 & 1.31 &  0.89 & 0.91 & 0.94 & 0.89 & 0.89 & -1.09 & -1.09 & 1.12 \\
-1.5 & 2 & 1.12 & 1.02 & 1.66 & 1.12 & 1.08 & 0.82 & 0.87 & 0.94 & 0.83 & 0.81 & -0.79& -0.81 & 1.18 \\      
-1.5 & 3& 0.98 & 0.83 & 1.66 & 0.99 & 0.94 & 0.76 & 0.81 & 0.94 & 0.78 & 0.74 & -0.57 & -0.64 & 1.21 \\
-1.5 & 5 & 0.83 & 0.62 & 1.66 & 0.88 & 0.78 & 0.66 & 0.70 & 0.94 & 0.69 & 0.62 & -0.29 & -0.41 & 1.30 \\
-1.0 & 3 & 0.64 & 0.48 & 1.20 & 0.69 & 0.62 & 0.57 & 0.55 & 0.90 & 0.55 & 0.53 & -0.07 & -0.09 & 1.31 \\
-0.5 & 3 & 0.42 & 0.29 & 0.78 & 0.44 & 0.41 & 0.33 & 0.24 & 0.79 & 0.31 & 0.31 & 0.43 &  0.36 & 1.22 \\
 0.0 & 3 & 0.29 & 0.21 & 0.44 & 0.30 & 0.29 & 0.18 & 0.11 & 0.50 & 0.17 & 0.17 & 0.93 & 0.76 & 1.15 \\
10 & 3 & .043 & .040& .043 & .044 & .043 & .004 & .003 & .004 & .004 & .004 & 10.9 & 10.1 & 1.00\\
\hline
\hline
\end{tabular}
\end{table}
\begin{table}
\caption{\label{table:5.3}Same as for Table~\ref{table:5.2} but for calculating $\chi$.
}
\begin{tabular}{cccccccccc}
\hline
\hline
 &  &  \multicolumn{5}{c}{$\chi_c$} & & &  \\
$\varepsilon_d$    & $U_{sd}$  & Ex. & Ren. & Unre. & Fit & XAS &  $\varepsilon_d^{\rm calc}$ & $\varepsilon_d^{\rm fit}$ & $t_{\rm eff}^{\rm fit}$\\
\hline
-1.5 & 1 &   0.12 & 0.10 & 0.05 & 0.12 & 0.13 & -1.09 & -1.09 & 1.12 \\
-1.5 & 2 &  0.20 & 0.19 & 0.05 & 0.20 & 0.23 & -0.79& -0.81 & 1.18 \\      
-1.5 & 3&   0.27 & 0.30 & 0.05 & 0.28 & 0.32 &-0.57 & -0.64 & 1.21 \\
-1.5 & 5 &  0.36 & 0.55 & 0.05 & 0.38 & 0.40 & -0.29 & -0.41 & 1.30 \\
-1.0 & 3 &  0.47 & 0.74 & 0.12 & 0.50 & 0.47 & -0.07 & -0.09 & 1.31 \\
-0.5 & 3 &  0.41 & 0.41 & 0.35 & 0.43 & 0.37 & 0.43 &  0.36 & 1.22 \\
 0.0 & 3 &  0.21 & 0.14 & 0.75 & 0.22 & 0.19 & 0.93 & 0.76 & 1.15 \\
10 & 3 &  .0006 & .0005& .0006& .0006& .0006& 10.9 & 10.1 & 1.00\\
\hline
\hline
\end{tabular}
\end{table}

We compare the exact results with several approximations \cite{xas}. In all these 
calculations $U_{sd}$ was put to zero and its effects were approximately 
included via renormalized parameter. The column ``Ren.'' shows results where
$\varepsilon_d$ was replaced by $\varepsilon_d^{\rm calc}=\tilde E_0(1)-\tilde E_0(0)$.
Here $\tilde E_0(n_d)$ is the energy of the model as a function of the occupancy $n_d$.
In this calculation the hopping to the localized level was cut to avoid double-counting. 
The Table also shows results for unrenormalized parameter (``Unre.''). We then
treated the $\varepsilon_d^{\rm fit}$ and $t_{\rm eff}^{\rm fit}$ as fitting parameters,
and adjusted these parameters to obtain the best possible agreement (``Fit'') 
with the exact results. Finally we have performed calculations where the hopping
matrix element to a level $\varepsilon_k$ 
\begin{equation}\label{eq:5.5}
[t_{\rm eff}(\varepsilon_k)]^2=t^2S(|\varepsilon_k-\varepsilon_F+\omega_0|),
\end{equation}
was related to the X-ray absorption or emission spectra. $[t_{\rm eff}(\varepsilon_k)]^2$
summed over all states is unrenormalized, but the hopping parameters to states 
close to the Fermi energy, $\varepsilon_F$ are enhanced at the cost of hopping
to the band edges. In calculations with $[t_{\rm eff}(\varepsilon_k)]$ we 
used the renormalized level position $\varepsilon_d^{\rm calc}$.

We first compare the exact results with the unrenormalized and renormalized results.
The renormalization improves the agreement with the exact results substantially. 
For most parameter sets  the agreement is relatively good. For $U_{sd}$ large
and for $|\varepsilon_d|$ rather small, there are still substantial deviations.
``XAS'' shows the results when the hopping is renormalized as well, using 
Eq.~(\ref{eq:5.5}). There is then a substantial additional improvement, and the
agreement is now generally a rather good. Finally, we have treated both the hopping 
and the level position as adjustable parameters. This gives only a marginal
improvement and sometimes the results are even worse. This is remarkable, since
the d-level position is now also a fit parameter and $\varepsilon_d^{\rm calc}$
is sometimes rather different from $\varepsilon_d^{\rm fit}$. On the other hand,
hopping is energy-dependent, and ``XAS'' presumably describes this better than ``Fit''.
This suggests that Eq.~(\ref{eq:5.5}) gives a quite good renormalization.  

In the case of Ce the delocalized states are primarily of $5d$ character. 
According to the Friedel sum rule we can then estimate the phase shift as $\delta\sim \pi/10$.
This then gives a singularity index of the order $\alpha \sim 0.1$. For thermodynamic
properties we may then expect an enhancement of the order of $(\tilde \omega/T_K)^{0.1}$,
where $T_K$ is the Kondo temperature.  For, e.g., CeCu$_2$Si$_2$ $T_K=0.001$ eV
and $t^2$ may then be enhanced by a factor of two, if we assume $\tilde \omega \sim $ 1 eV. 
As discussed above, we do not advocate including these effects explicitly in a model. 
However, one should be aware that thermodynamic and spectroscopic properties may be 
renormalized differently.

\section{Fullerenes}\label{sec:6}

In this section we discuss the parameters for a molecular solid.
As an example we use fullerenes \cite{book}\index{fullerenes}. Similar work has been done 
for TTF-TCNQ \cite{Erik}.

\subsection{Hopping}\label{sec:6.1}

\begin{figure}[h]
 \centering
 \includegraphics[angle=-90,width=0.4\textwidth]{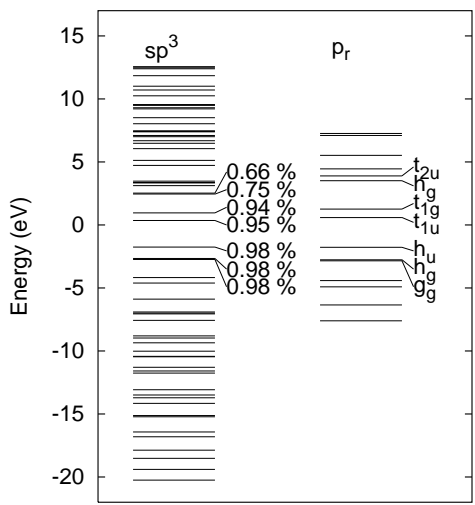}       
 \caption{\label{fig:6.1}Levels of the C$_{60}$ molecule. The
left-hand part shows the levels obtained by using a basis
of one $2s$ and three $2p$ orbitals per carbon atom ($sp^3$). The
right-hand part shows the levels obtained by using just one
radial $2p$ orbital per atom ($2p_r$). The numbers give the
amount of radial $2p$ character ($2p_r$) in the full calculation
(after Ref. \cite{book}).}
\end{figure}
The important levels in a C$_{60}$ molecule can be described in a 
tight-binding picture including a $2s$ and three $2p$ orbitals on 
each of the 60 C atoms.  The corresponding molecular levels are 
shown in Fig.~\ref{fig:6.1}. The molecule forms approximate $sp^2$ 
hybrids on each C atom which point towards the neighboring 
C atoms and radial orbitals $p_r$ pointing out of the molecule. 
The former orbitals interact strongly and form bonding and antibonding 
molecular orbitals at the lower and upper end of the spectrum, 
respectively. The $p_r$ orbitals interact much less and form molecular 
orbitals in the middle of the spectrum. The figure illustrates that 
these orbitals can be described rather well using only the $p_r$ orbitals. 
In the neutral molecule all orbitals up to and including the $h_u$ orbital 
are filled.

C$_{60}$ molecules condense to a solid of rather weakly bound molecules.
Thus the distance ($\sim 3$ \AA) between the closest C atoms on two neighboring 
molecules is much larger than the distance ($\sim 1.4$ \AA) between two C atoms on the
same C$_{60}$ molecule. The molecular levels then essentially preserve their identity
in the solid, but the discrete molecular levels are broadened to narrow essentially
nonoverlapping band. The alkali-doped fullerenes are of particular interest.
In these systems the $t_{1u}$ band is partly filled. Therefore the three-fold degenerate 
$t_{1u}$ molecular level is particularly interesting. 

\begin{figure}[h]
 \centering
 \includegraphics[width=0.5\textwidth]{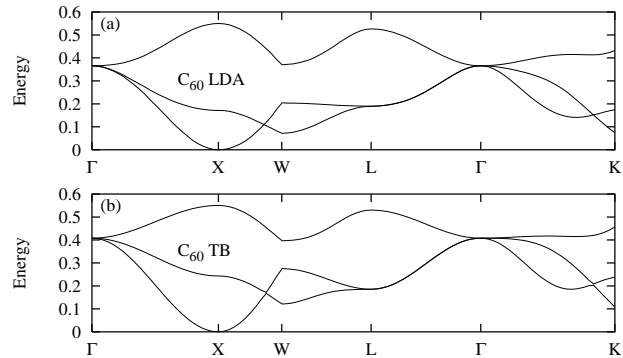}
 \caption{\label{fig:6.2}Band structure for a C$_{60}$ solid in
the Fm${\bar 3}$ structure (a) according to an {\it ab initio} LDA and (b) according to
a TB calculation (after Gunnarsson {\it et al.}\cite{alkalia4}).
}
\end{figure}

The band structure can be described in a tight-binding\index{tight-binding} 
(TB) scheme. We first form a molecular orbital corresponding to the $t_{1u}$ 
level. The hopping between the molecules is described by hopping integrals 
$V_{pp\sigma}$ and $V_{pp\pi}$\index{hopping integrals} 
corresponding to hopping between orbitals pointing directly towards each other 
or orbitals pointing perpendicular to the connecting line of the centers.
Following Harrison,\cite{Harrison} we assume that the ratio
of the $\pi-$ and $\sigma$-integrals is -1/4. Then
\begin{equation}\label{eq:hopp2c}
V_{pp\sigma}=v_{\sigma}{R\over R_0}e^{-\lambda(R-R_0)}; \hskip0.5cm 
{V_{pp\pi} \over V_{pp\sigma}}=-{1\over 4}\hskip0.5cm R_0=3.1 
\ {\rm \AA},
\end{equation}
where $R$ is the separation of the carbon atoms. The prefactor $R$
has been included to simulate the $r$-dependence of a $2p$ orbital
as described by Slater's rules\cite{Slater}. The overall hopping strength, 
determined by $v_{\sigma}$, is adjusted to the band width in a band 
structure calculation, and the decay length $\lambda$ is determined 
from the dependence of the band width on the lattice parameter.
Here we use the parameters \cite{c60tb,alkalia4}
\begin{equation}\label{eq:hopp2d}
\lambda=1.98 \ {\rm \AA}^{-1} {\rm and} \hskip0.3cm v_{\sigma}=0.917
\ {\rm eV}.                
\end{equation}
The resulting TB band structure is compared with an {\it ab initio}
band structure calculation in Fig.~\ref{fig:6.2}. The agreement is quite good. 
The resulting band structure $\varepsilon_{\bf k}$ has a simple parameterization
\cite{c60tb,Nozha}. The dominating hopping between two molecules in this structure is 
given by two equivalent hopping integrals, with all other hopping integrals being 
substantially smaller. Effectively, we have therefore adjusted this parameter
requiring that the TB band width should agree with the LDA band width. The shape of
the band structure in Fig. \ref{fig:6.2} is therefore primarily
determined by the geometry of the C$_{60}$ molecule and by the relative
positions and orientations of the C$_{60}$ molecules in the Fm${\bar 3}$
symmetry. 

\subsection{Coulomb interaction}\label{sec:6.2}

We first consider the Coulomb integral\index{Coulomb integral} $U_0$ between two $t_{1u}$ electrons
for a free C$_{60}$ molecule.
A very simple estimate is obtained by assuming that charge density of the $t_{1u}$ orbital
forms a thin shell of charge on a sphere  with the radius
$R\sim 3.5 $ \AA. Then $U_0=e^2/R=4$ eV. This neglects that the orbitals breath when the occupancy
is changed. To obtain a better estimate one can calculate how the $t_{1u}$ eigenvalue 
changes with occupancy, using Eq.~(\ref{eq:4.5}) and LDA. This leads to values of
the order 2.7-3.0 eV \cite{Martins1,Pederson,AntropovU}. $U_0$ for a free molecule 
can also be estimated from experimental results using 
\begin{equation}\label{eq:6.1}
U_0= I_p(C_{60}^{-})-A(C_{60}^{-})=E_0(2)+E_0(0)-2E_0(1),
\end{equation}
where $E_0(n)$ is the energy of a C$_{60}$ molecule with $n$ $t_{1u}$ electrons.
This leads to $U_0\sim 2.7$ eV \cite{book,A}, in fairly good agreement with theory.

We next consider $U$ for a C$_{60}$ solid, following Antropov {\it et al.}\cite{AntropovU}. 
$U$ is strongly screened by the polarization of the surrounding molecules. To 
describe this we put the C$_{60}$ molecules on an fcc lattice and assign a 
polarizability $\alpha$ to each molecule. An electron is added to the central 
molecule, and the surrounding molecules are allowed polarize in a
self-consistent way. This polarization acts back on the electron and 
reduces the energy increase of the $t_{1u}$ level by an amount $\delta U$.  
The summation over neighboring molecules is extended until it is converged. The
$U$ for the solid is then
\begin{equation}\label{eq:6.2}       
U=U_0-\delta U. 
\end{equation}
The value of $\alpha$ can be determined from the experimental
value of the dielectric function (4.4).\cite{eps} Using the
Clausius-Mossotti relation and  the lattice parameter $a=14.04$
\AA, this leads to $\alpha=90$ \AA$^3$.  {\it Ab initio} calculations
using the density functional formalism gave $\alpha=83$ \AA$^3$\cite{Pederson}. 
Using $\alpha=90$ \AA$^3$, Antropov {\it et al.}\cite{AntropovU} found 
$\delta U=1.7$ eV.  Together with $U_0=2.7$, this gives $U=1.0$ eV.  These 
values of $U$ do not include the metallic screening\index{screening} from the
$t_{1u}$ electrons in A$_3$C$_{60}$ compounds, and they are appropriate
for models where the metallic screening is treated explicitly when
solving the corresponding model.

We next consider the nearest neighbor interaction $V$, which
is obtained by calculating the increase of the energy of a
$t_{1u}$ orbital on a molecule $1$ when an electron is added
to a neighboring molecule $2$. This leads to the result
\begin{equation}\label{eq:6.3}    
V=e^2/R-\delta V,
\end{equation}
where $R$ is the nearest neighbor separation and $-\delta V$
is the lowering of the $t_{1u}$ orbital on molecule $1$ due to
the polarization of the surrounding molecules when an electron
is added to molecule $2$. For $a=14.04$ \AA, Antropov
{\it et al.}\cite{AntropovU} estimated that $\delta V$=1.12 eV,
resulting in $V=0.3$  for the polarizability $\alpha$=90 \AA$^3$.
The same value $V=0.3$ eV was also obtained by Pederson and 
Quong\cite{Pederson}. We can see that $U$ is indeed substantially 
larger than $V$, and that it is justified to focus on the effects 
of $U$ at first.

$U$ can be  estimated experimentally from Auger
spectroscopy\cite{Sawatzky,Martensson}. A carbon $1s$
electron is emitted in a photoemission process. This is followed
by an Auger process, where a carbon $2p$ electron falls down into
the $1s$ hole and another $2p$ electron is emitted. For noninteracting 
electrons, the Auger spectrum is just the self-convolution of the 
photoemission spectrum.  For the interacting system, the Auger spectrum 
is expected to be shifted due to the interaction of the two holes in 
the final state. Indeed, Lof {\it et al.}\cite{Sawatzky} found good
agreement with the self-convoluted curve when this was shifted
by 1.6 eV. The experimental estimate of the Coulomb interaction
is then $U=1.6\pm 0.2$ eV\cite{Sawatzky} as an average over all orbitals
and about 1.4 eV for the highest occupied orbital. Since Auger is rather 
surface sensitive, this number may be more representative for $U$ 
at the surface.  One can estimate that $U$ at the surface is about 
0.3 eV larger than in the bulk, due to fewer neighbors and less efficient
screening\cite{AntropovU}.  This suggests that the bulk value
of $U$ for the $t_{1u}$ and $h_u$ orbitals may be on the order
$U=1.1$ eV, which close to the theoretical estimate.         
$U$ has also been estimated for K$_6$C$_{60}$
in a similar way\cite{Martensson}, and giving a similar value
$U=1.5$ eV.

\subsection{Electron-phonon interaction}\label{sec:6.3}

\begin{figure}[h]
 \centering
 \includegraphics[angle=-90,width=0.5\textwidth]{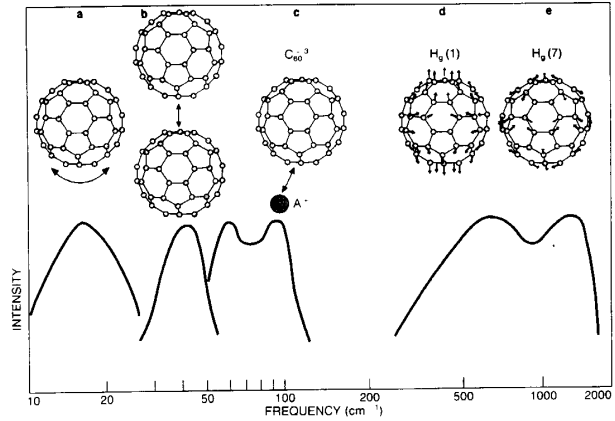}
\caption[]{\label{fig:6.3}
Schematic representation of various phonons in A$_3$C$_{60}$
compounds. The figure shows, from left to right,
(a) librations, (b) intermolecular C$_{60}$-C$_{60}$ phonons,
(c) A-C$_{60}$ phonons and (d)-(e) intramolecular H$_{g}$ modes.
The figure indicates the radial and tangential character of the
low-lying and and high-lying H$_g$ modes, respectively (After
Hebard\cite{Hebard}).}
\end{figure}

The electron-phonon interaction\index{electron-phonon interaction} 
plays an important role for many properties of alkali-doped fullerides. 
For instance, superconductivity is believed to be due the electron-phonon
interaction. Fig. \ref{fig:6.3} indicates the
different types of phonons in alkali-doped C$_{60}$ compounds.
The low-lying modes are librations (4-5 meV) and intermolecular modes 
(energies up to about 17 meV) involving alkali-C$_{60}$ and C$_{60}$-C$_{60}$ 
modes. The high-lying modes (34-195 meV) are intramolecular modes,
where the molecules are deformed. All the low-lying modes have a 
rather weak coupling to the $t_{1u}$ electrons, and the main coupling 
is to the intramolecular phonons. Here, we therefore focus on the the 
coupling to these phonons. These phonons couple primarily to the level
energies  in contrast to the intermolecular phonons which couple to the 
hopping integrals.

The C$_{60}$ molecule has $60\times 3-6=174$ intramolecular modes. For 
symmetry reasons, however, the $t_{1u}$ electrons only couple to modes with
A$_g$ or H$_g$ symmetry. There are eight five-fold degenerate H$_g$
modes and two nondegenerate A$_g$ modes. The coupling to the $t_{1u}$ level 
takes the form\cite{C60jt}          
\begin{eqnarray}\label{eq:6.4}
&&H_{el-ph}=   
\sum_{\nu=1}^8g_{\nu}\sum_{M=1}^5\sum_{\sigma}
\sum_{m^{\phantom '}=1}^3\sum_{m^{'}=1}^3 \lbrack V_{H_g}^{(M)}
\rbrack_{m^{\phantom '}m^{'}}
\psi_{m\sigma}^{\dagger}
\psi_{m^{'}\sigma}(b_{\nu M}^{\phantom \dagger}+b_{\nu M}^{\dagger}) \cr
&&+\sum_{\nu=9}^{10} g_{\nu} \sum_{\sigma}
\sum_{m^{\phantom '}=1}^3\sum_{m^{'}=1}^3 \lbrack V_{A_g}
\rbrack_{m^{\phantom '}m^{'}}\psi_{m\sigma}^ {\dagger} 
\psi_{m^{'}\sigma} (b_{\nu}^{\phantom \dagger}+b_{\nu}^{\dagger}),
\end{eqnarray}
where $\psi_{m\sigma}^{\dagger}$ creates a $t_{1u}$ electron with quantum number $m$
and $b_{\nu M}^{\dagger}$ creates a phonon in mode $\nu$ with quantum number $M$.
The first eight modes are H$_g$ Jahn-Teller phonons\index{Jahn-Teller phonons} and the next two A$_g$ phonons. The coupling
constants are $g_{\nu}$ and 
the coupling to the H$_g$ phonons is given by the matrices
\begin{eqnarray}\label{eq:6.5}  
&&V_{H_g}^{(1)}=
 {1\over 2}\left(\begin{array}{ccc}
   -1 & 0 & 0  \\                                      
   0  &-1 & 0  \\                                      
   0  & 0 & 2  \\                                      
\end{array}\right)\ \hskip0.3cm 
V_{H_g}^{(2)}={\sqrt{3}\over 2}
 \left(\begin{array}{ccc}
   1  & 0 & 0  \\                                      
   0  & -1& 0  \\                                      
   0  & 0 & 0  \\                                      
\end{array}\right)\ \hskip0.3cm
V_{H_g}^{(3)}={\sqrt{3}\over 2}
 \left(\begin{array}{ccc}
   0  & 1 & 0  \\                                      
   1  & 0 & 0  \\                                      
   0  & 0 & 0  \\                                      
\end{array}\right)\ \cr         
&&V_{H_g}^{(4)}={\sqrt{3}\over 2}
 \left(\begin{array}{ccc}
   0  & 0 & 1  \\                                      
   0  & 0 & 0  \\                                      
   1  & 0 & 0  \\                                      
\end{array}\right)\    
\hskip0.7cm V_{H_g}^{(5)}={\sqrt{3}\over 2}
 \left(\begin{array}{ccc}
   0  & 0 & 0  \\                                      
   0  & 0 & 1  \\                                      
   0  & 1 & 0  \\                                      
\end{array}\right)\    
\end{eqnarray}
and the coupling to the A$_g$ phonons by
\begin{equation}\label{eq:6.6}       
V_{A_g}^{(1)}=
 \left(\begin{array}{ccc}
    1 & 0 & 0  \\                                      
   0  & 1 & 0  \\                                      
   0  & 0 & 1  \\                                      
\end{array}\right)\ . 
\end{equation}
The corresponding dimensionless electron-phonon coupling constant
is\cite{C60jt}        
\begin{equation}\label{eq:6.7}
\lambda ={5\over 3}N(0)\sum_{\nu=1}^8{g_{\nu}^2\over \hbar \omega_{\nu}}
+{2\over 3}N(0)\sum_{\nu=9}^{10}{g_{\nu}^2\over \hbar \omega_{\nu}},
\end{equation}
where $N(0)$ is the density of states per spin and molecule
and $\omega_{\nu}$ is the frequency of the mode $\nu$.
 
The theoretical calculation of the electron-phonon coupling for a solid is
very complicated. Lannoo {\it et al.}\cite{Lannoo} showed that for intramolecular
modes in fullerides, important simplifications follow from the large
difference between the intramolecular ($E_I$) and intermolecular ($W$)
energy scales. The coupling for a solid can then be obtained approximately 
from a calculation for a free molecule and the density of states $N(0)$ of the solid.
Thus, it is sufficient to calculate the shift $\Delta \varepsilon_{\nu 
\alpha}$  of the $t_{1u}$ levels $\alpha$ for a free C$_{60}$ molecule
per unit displacement of the $\nu$th phonon coordinate. One then finds
that
\begin{equation}\label{eq:6.8}
\lambda \sim N(0)\sum_{\nu\alpha}{\Delta \varepsilon_{\nu\alpha}^2
\over \omega_{\nu}^2}.
\end{equation}
This gives a molecular specific quantity which is multiplied by $N(0)$.
Table~\ref{table:6.1} shows results for the electron-phonon coupling.
The theoretical calculations by Antropov {\it et al} \cite{Antropovphon}, 
Faulhaber {\it et al.}\cite{Faulhaber} and Manini {\it et al.}\cite{Manini}
are based on {\it ab initio} LDA calculations. The work of Iwahara {\it et al.} is
based on the B3LYP functional with some Hartree-Fock exchange mixed in.
There are substantial deviations between the distribution of coupling strength
to the different modes in the different calculations. This distribution
is very sensitive to the precise form of the phonon eigenvectors.
The deviations between the total coupling strengths are smaller. The work of 
Iwahara {\it et al.} gives a stronger coupling than the other three calculations.
This is not so surprising, since this work is based on a rather different 
functional.

\begin{table}
\caption[]{\label{table:6.1}Partial electron-phonon coupling
constants $\lambda_{\nu}/N(0)$ (in eV) according to different 
theoretical calculations and derived from photoemission and 
Raman scattering.  The energies $\omega_{\nu}$ (in cm$^{-1}$) of
the modes for the undoped system are shown.}
\vskip0.2cm
\begin{tabular}{ccccccccc}
\hline
\hline
  &  &  \multicolumn{7}{c}{$\lambda_{\nu}/N(0)$}  \\
  &   & \multicolumn{4}{c}{Theory} & \multicolumn{2}{c}{Photoemission} & Raman\\
Mode & $\omega_{\nu}$   & Antrop. \cite{Antropovphon}& Faul.\cite{Faulhaber} &
Man.\cite{Manini}  & Iwa.\cite{Iwahara} &   Gun.\cite{Gunnarssonpes}       & Iwa.\cite{Iwahara} & Kuz.\cite{Winter}    \\
\hline
H$_g$(8)&1575& .022&  .009 & .014 & .018  & .023   &  .011 & .003 \\
H$_g$(7)&1428& .020 & .015 & .015 & .023  & .017   &  .028 & .004   \\
H$_g$(6)&1250& .008 & .002 & .003 & .002  & .005   &  .007 & .001   \\
H$_g$(5)&1099& .003 & .002 & .004 & .005  & .012   &  .009 & .001   \\
H$_g$(4)& 774& .003 & .010 & .004 & .006  & .018   &  .007 & .003   \\
H$_g$(3)& 710& .003 & .001 & .009 & .012  & .013   &  .015 & .003   \\
H$_g$(2)& 437& .006 & .010 & .011 & .011  & .040   &  .012 & .020   \\
H$_g$(1)& 273& .003 & .001 & .005 & .006  & .019   &  .007 & .048   \\
\hline
$\sum$ H$_g$ &  & .068 & .049 & .065 & .083  & .147   &  .096 & .083   \\
\hline
\hline
\end{tabular}
\label{table:pa2}
\end{table}

An experimental method for estimating the electron-phonon coupling is 
the use of photoemission data. Because of the relatively strong electron-phonon
coupling, we expect to see satellites due to the excitation of phonons.
The weights of the satellites give information about the strength of
the coupling. This is essentially the Franck-Condon effect, but because of 
the Jahn-Teller effect the calculation of the satellite structure is rather 
complicated. The photoemission spectra  of K$_3$C$_{60}$ and Rb$_3$C$_{60}$ have
been analyzed along these lines\cite{Knupfer93}. Due to the broadening
effects in a solid and due to the complications in the theoretical
treatment of bands with dispersion, however, it was not possible to
derive reliable, quantitative values for the electron-phonon coupling.

\begin{figure}[h]
 \centering
 \includegraphics[angle=-90,width=0.6\textwidth]{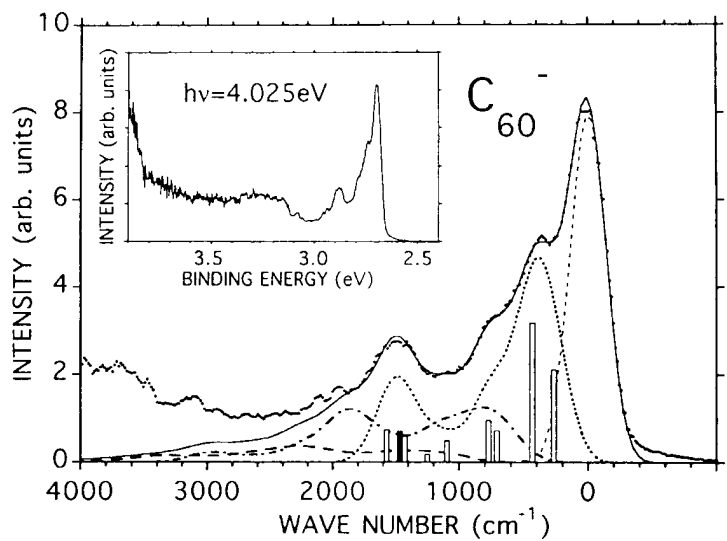}       
\caption[] {\label{fig:6.4}Experimental (dots) and theoretical (full line)
photoemission spectrum of C$_{60}^-$. The theoretical no loss (dashed),
single loss (dotted) and double loss (dashed-dotted) curves are also
shown. The contributions of the different modes to the single loss curve
are given by bars (H$_g$: open, A$_g$: solid).
The inset shows the experimental spectrum over a larger energy range
(after Gunnarsson {\it et al.}\cite{Gunnarssonpes}).}
\end{figure}

Photoemission\index{photoemission} spectra have also been measured for free C$_{60}^{-}$
molecules. In this case the theoretical treatment is substantially
simpler\cite{Gunnarssonpes}. In these experiments, a beam of C$_{60}^{-}$
ions was created and a photoemission experiment was performed using
a laser light source ($\hbar \omega$ =4.025 eV)  and a time of flight
spectrometer. The spectrum resulting from emission from the $t_{1u}$ level
was measured. To analyze the results, we use the coupling in Eq.~(\ref{eq:6.4}) 
of the $t_{1u}$ level to the two A$_g$ and the eight five-fold degenerate 
H$_g$ modes. For this model the ground-state can be calculated by numerical 
diagonalization to any desired accuracy\cite{Gunnarssonpes}. Furthermore, 
within the sudden approximation\cite{Hedin69}, the photoemission spectrum
can  easily be calculated. A set of coupling constants are then assumed
and the resulting spectrum is compared with experiment. The coupling
parameters are varied until good agreement with experiment is obtained,
thereby providing an estimate of the couplings.  The resulting spectrum is
compared with experiment in Fig. \ref{fig:6.4} and the corresponding
parameters are shown in Table \ref{table:6.1}. An uncertainty in this 
approach is that with the available resolution,  it is not possible to 
distinguish between the coupling to A$_g$ modes and H$_g$ modes with similar 
energies. The couplings to the A$_g$ modes were therefore taken from a
calculation\cite{Antropovphon}. With this assumption, the couplings
to the H$_g$ modes can then be determined. An equally good fit can, however,
be obtained using other couplings to the A$_g$ modes if the couplings to
the H$_g$ modes are changed correspondingly. 

Substantially later the experiments in Ref. \cite{Gunnarssonpes} were repeated
by Wang {\it et al.}\cite{Wang}. It was now possible to obtain a better resolution. 
These data have been analyzed in a similar way as in Ref. \cite{Gunnarssonpes}
by Iwahara {\it et al.}\cite{Iwahara}. There results are shown in Table~\ref{table:6.1}.
The total coupling is weaker than in Ref. \cite{Gunnarssonpes}, but still 
substantially larger than in the {\it ab initio} LDA calculations.
The agreement with the calculation using the B3LYP is better.

Raman scattering\index{Raman scattering} provides a different method of estimating the coupling strength.
The electron-phonon coupling allows phonons to decay into an electron-hole pair
in the metallic fullerides. This decay contributes to the width of the phonon 
and can be measured in Raman scattering. Other factors may also contribute to 
the width, but one can try to eliminate these by subtracting the width of the
phonons for a nonmetallic system, where a decay in electron-hole pairs is not
possible. This was done by Winter and Kuzmany\cite{Winter}, and Table~\ref{table:6.1} 
shows results adapted \cite{book} from the experiments \cite{Winter}.
The total weight does not differ much from what was obtained from photoemission\cite{Iwahara} ,
but the distribution of weight between the different modes differs dramatically. 
Theoretically, it is found that in the solid there is a transfer of weight
to lower modes, due to the coupling to electron-hole pairs \cite{jongcoul}.
This mechanism is operative for the Raman data but not for the photoemission data 
(taken for a free molecule). This may explain some of the discrepancy between 
the PES and Raman data.

\section{Conclusions}

For complicated systems with strong correlation effects it is often
not possible to obtain accurate {\it ab initio} solutions, but it 
is instead useful to turn to models. An important issue is then
how to obtain parameters and how to renormalize parameters to include
as much physics as possible. We have discussed how the basic principle 
is to try to include implicitly as a renormalization of parameters all
effects not explicitly included in the model. On the other hand, we should 
not allow effects included explicitly in the model to renormalize parameters. 
For many-body systems there is no general systematic and controlled way of
doing this. The basic assumption is often that the electrons can be put into
two groups of ``fast'' (delocalized) and ``slow'' (localized) electrons, 
where the ''fast'' electrons are assumed to adjust to the ``slow'' electrons,
and therefore can projected out. Such a division is, however, often not very 
clear cut.  Nevertheless some methods have been relatively successful
in obtaining parameters for certain classes of systems. We have, however, 
shown simple examples of many-body effects that are usually not included,
but can have an appreciable effect on the parameters. In particular, renormalization effects 
may work differently for different experiment. We have also argued that it is 
important to try to extract parameters from different sources, both theory and 
experiment, to obtain a better understanding of  the accuracy of the parameters.  



\begin{thebibliography}{99}
\bibitem{note}Ref. \cite{Ferdi} (Sec.~IIID) argues that the ``cut off'' LDA\cite{McMahan,cut} for, e.g.,
a transition metal, incorrectly includes on-site $3d-3d$ transitions to describe 
breathing of the central $3d$ orbitals. This is incorrect, since $3d-3d$ transitions would 
not change the radial extent of the $3d$ orbital. It was further implied that metallic 
$3d$ screening is included, i.e., transfer of charge to the central $3d$ orbitals,
already included explicitly in the Hubbard model. 
This is, however, excluded by keeping the total occupancy of the $3d$ orbitals 
at the prescribed value, and there is no double-counting. Finally, it was argued 
that breathing of the $3d$ orbital should not be included. Since breathing is not 
explicitly included in the Hubbard model, it should  be included in the calculation 
of $U$. It is an important effect \cite{Mn,config}.
\bibitem{Drchal}
Similar results were found for other system containing atoms from the middle of the $3d$ series\cite{Drchal1}.
\end{thebibliography}
\end{document}